\documentclass[suppldata]{gOMS2e}

\usepackage{epstopdf}
\usepackage{subcaption}
\usepackage{lipsum}
\usepackage{amsmath}
\usepackage{mathtools}
\usepackage[usenames,dvipsnames]{color}
\usepackage{hyperref}
\usepackage{multirow}
\usepackage{siunitx}
\theoremstyle{plain}

\theoremstyle{definition}

\theoremstyle{remark}

\newcommand{\real}{{\mathbb R}}

\newcommand{\grey}[1]{\textcolor{Gray}{#1}}
\newcommand{\changecol}[1]{#1}
\newcommand{\changecolnew}[1]{#1}

\makeatletter
\usepackage{xspace}
\def\@onedot{\ifx\@let@token.\else.\null\fi\xspace}
\DeclareRobustCommand\onedot{\futurelet\@let@token\@onedot}

\newcommand{\figref}[1]{Fig\onedot~\ref{#1}}
\newcommand{\equref}[1]{Eq\onedot~\eqref{#1}}
\newcommand{\secref}[1]{Sec\onedot~\ref{#1}}
\newcommand{\tabref}[1]{Tab\onedot~\ref{#1}}

\newcommand{\logsumexp}{\operatornamewithlimits{logsumexp}}

\def\bmat {\begin{bmatrix}}
\def\emat {\end{bmatrix}}

\usepackage{bm}
\newcommand{\vectorstyle}[1]{\bm{#1}}
\newcommand{\matrixstyle}[1]{{\bf #1}}
\newcommand{\tensorstyle}[1]{\bm{\mathsf #1}}

\newcommand{\vp}{\vectorstyle{p}}
\newcommand{\vc}{\vectorstyle{c}}
\newcommand{\ve}{\vectorstyle{e}}
\newcommand{\vm}{\vectorstyle{m}}
\newcommand{\vr}{\vectorstyle{r}}
\newcommand{\vl}{\vectorstyle{l}}
\newcommand{\vu}{\vectorstyle{u}}
\newcommand{\vq}{\vectorstyle{q}}
\newcommand{\vv}{\vectorstyle{v}}
\newcommand{\vx}{\vectorstyle{x}}
\newcommand{\vt}{\vectorstyle{t}}
\newcommand{\vy}{\vectorstyle{y}}
\newcommand{\vz}{\vectorstyle{z}}
\newcommand{\vw}{\vectorstyle{w}}
\newcommand{\valpha}{\vectorstyle{\alpha}}
\newcommand{\vkappa}{\vectorstyle{\kappa}}
\newcommand{\vmu}{\vectorstyle{\mu}}

\newcommand{\mQ}{\matrixstyle{Q}}
\newcommand{\mL}{\matrixstyle{L}}
\newcommand{\mM}{\matrixstyle{M}}
\newcommand{\mX}{\matrixstyle{X}}
\newcommand{\mY}{\matrixstyle{Y}}
\newcommand{\mZ}{\matrixstyle{Z}}
\newcommand{\mV}{\matrixstyle{V}}
\newcommand{\mW}{\matrixstyle{W}}
\newcommand{\mU}{\matrixstyle{U}}
\newcommand{\mR}{\matrixstyle{R}}
\newcommand{\mE}{\matrixstyle{E}}
\newcommand{\mT}{\matrixstyle{T}}
\newcommand{\mMu}{\matrixstyle{M}}
\newcommand{\mSigma}{\matrixstyle{\Sigma}}

\newcommand{\tM}{\tensorstyle{M}}
\newcommand{\tT}{\tensorstyle{T}}

\newcommand{\tSigma}{\tensorstyle{\Sigma}}

\newcommand{\transpose}{\top}

\def\eg{\emph{e.g}\onedot} 
\def\ie{\emph{i.e}\onedot} 
\def\etc{\emph{etc}\onedot} 
 
\def\etal{\emph{et al}\onedot}

\newcommand{\Cpp}{C\texttt{++}}

\begin{document}

\jvol{00} 
\jnum{00} 
\jyear{2018} 
\jmonth{Feb}

\pagestyle{myheadings}
\markright{\v Srajer, Kukelova, Fitzgibbon\hfill A Benchmark of AD Tools \hfill}


\title{\vspace*{-24mm}
A Benchmark of Selected Algorithmic Differentiation Tools on \\
Some Problems in Computer Vision and Machine Learning\thanks
{Shorter versions of this article appeared at AD2016---7th International Conference on Algorithmic Differentiation, and in Optimization Methods and Software, Taylor and Francis, Feb 2018 (online).%
}}

\author{
\name{
Filip \v Srajer\textsuperscript{a} 
\qquad
Zuzana Kukelova\textsuperscript{b}$^{\ast}$%
\thanks{$^\ast$Corresponding authors. Email: kukelzuz@fel.cvut.cz, awf@fitzgibbon.ie} 
\qquad
Andrew Fitzgibbon\textsuperscript{c$\ast$}}
\affil{%
\textsuperscript{a}Department of Computer Science, ETH Zurich, Switzerland\\
\textsuperscript{b}Faculty of Electrical Engineering, CTU in Prague, Czech Republic \\
\textsuperscript{c}Microsoft, Cambridge, United Kingdom}
}

\maketitle

\begin{abstract}
Algorithmic differentiation (AD) allows exact computation of derivatives given only an implementation of an objective function.  Although many AD tools are available, a proper and efficient implementation of AD methods is not straightforward. The existing tools are often too different to allow for a general test suite. In this paper, we compare fifteen ways of computing derivatives including eleven automatic differentiation tools implementing various methods and written in various languages (\Cpp{}, F\#, MATLAB, Julia and Python), two symbolic differentiation tools, finite differences, and hand-derived computation.

We look at three objective functions from computer vision and machine learning. These objectives are for the most part simple, in the sense that no iterative loops are involved, and conditional statements are encapsulated in functions such as {\tt abs} or {\tt logsumexp}. However, it is important for the success of algorithmic differentiation that such `simple' objective functions are handled efficiently, as so many problems in computer vision and machine learning are of this form.

Of course, our results depend on programmer skill, and familiarity with the tools.  However, we contend that this paper presents an important datapoint: a skilled programmer devoting roughly a week to each tool produced the timings we present.  We have made our implementations available as open source to allow the community to replicate and update these benchmarks.
\end{abstract}


\begin{classcode}65D; 68T\end{classcode}

\section{Introduction}
\label{sec:intro}
Algorithmic differentiation (AD) is a set of methods for automatic and exact computation of derivatives given a definition, in source code, of a function to be differentiated.  It includes automatic differentiation, where derivatives are forward and/or back propagated through the chain rule. This is possible since even the most complicated functions are composed of elementary operations and functions such as addition, multiplication, logarithm, exponential, \etc.

Alternative approaches to automatic differentiation include symbolic differentiation, finite differences and differentiation by hand.  Symbolic differentiation using symbolic algebra systems typically has to represent the whole function as a single expression, which is limited by available memory, meaning it cannot handle larger functions. For efficient code generation, it should also include common subexpression elimination.  

Finite differences (FD) is a numerical method and therefore does not compute exact derivatives.  Note however that for most computer vision an machine learning problems, this inaccuracy is often unimportant~\cite{Triggs99ba}.
Of more importance is the computational cost: the asymptotic time complexity is dependent on the number of input variables whereas the complexity of so called reverse mode of AD is independent of it.

Finally, differentiating functions manually by a human is very time consuming and also error prone, but almost always results in the fastest runtime code.

As mentioned above, AD exploits the chain rule for computing derivatives. The chain rule is typically traversed either in the direction from the input variables to the output variables (forward mode) or the other way around (reverse mode).
Asymptotic time complexity of forward mode is dependent on the number of input variables and complexity of reverse mode on the number of output variables. Hence, a mode should be chosen based on a function to be differentiated. Note that there are also hybrid ways of computing derivatives using AD which are not precisely forward or reverse mode. For a more detailed explanation of AD methods, see Griewank and Walther~\cite{Griewank08} and Baydin \etal~\cite{Baydin15survey}.

Usually, AD is implemented by operator overloading (OO) or source transformation (ST). As an example, consider a \Cpp{} function working with floating point variables. An operator overloading tool requires that the function is written in terms of a templated type. Then, the tool instantiates the function template with a custom type which stores not only a variable but also a value of its derivative. This custom type overloads all elementary operations to also update the derivative value. Consequently, the output of the function includes the final value of the derivative.  This corresponds to the forward mode.  Reverse mode is sometimes considered more complicated, but the main idea is similar.  On the other hand, source transformation tools analyze the original function, somewhat as a compiler would, and output source code for a function which computes the derivative. Source transformation can potentially output a code computing derivatives more efficiently than operator overloading tools but it is usually much more difficult to implement as it has to know the syntax of the desired programming language.

Many AD tools exist (see \cite{autodiff-org} and \tabref{tab:all}). Nevertheless, it is not trivial to implement one properly, especially so that it could be used for complicated objective functions. The existing tools are in various languages and implement various AD methods. Hence, most of the tools are too different to allow for a straightforward implementation of benchmark suites. 

We propose to take three objective functions from machine learning and computer vision, to benchmark eleven selected AD tools covering various languages and AD methods (see \tabref{tab:all}), two symbolic differentiation tools, finite differences and also hand-derived derivative computation. The objective functions considered are: log-likelihood of a Gaussian mixture model, bundle adjustment~\cite{Triggs99ba}, and hand tracking~\cite{Taylor14hand}. These functions include features such as sparse Jacobians, matrix expressions, and domain-specific special functions such as logsumexp, defined stably as
\begin{equation}
\logsumexp(\vx : \mathbb R^n) = \log(\operatorname{sum}(\exp(\vx - \max(\vx)))) + \max(\vx)
\end{equation}

Recently, Siskind and Pearlmutter~\cite{Siskind16} presented a benchmark of several AD tools. They show runtimes relative to the runtime of their own tool whereas we give absolute runtimes as well as runtimes normalized with respect to individual languages (see \secref{sec:exp}). Their objective functions are simple with a fixed number of input and output variables. On the other hand, all our problems have varying number of variables.  D{\"u}rrbaum et al.~\cite{Durrbaum02comparison} benchmarked ADOLC versus symbolic differentiation and found significant speed differences, also borne out by our experiments.

We first give an overview of the AD tools selected for benchmarking. Next, we briefly present how AD is used in machine learning and computer vision followed by a description of objective functions used for benchmarking in this work. Then, we present the results and finally give our conclusions, foremost among which is that even with reasonable care devoted to efficiency in each of the input languages, the runtimes vary through four orders of magnitude.  While factors other than speed are important, it should always be kept in mind that for many applications, finite difference computation is sufficiently accurate, and it is certainly the easiest to use, so any tool, to be valuable, must beat FD for speed.

\subsection{Notation}
\label{sec:not}
In this paper, we use the following notation for variables: scalar $s$ or $S$, vector $\vv$, matrix $\mM$, and tensor $\tT$. We symbolize a concatenation of multiple column vectors $\vv_1, \vv_2, \dots, \vv_n$ as a matrix $\mV$. Similarly, a concatenation of multiple matrices $\mM_1, \mM_2, \dots, \mM_m$ as a tensor $\tM$.

Special functions are matrix determinant or scalar absolute value $|\cdot|$, and Euclidean norm $||\cdot||$. Function logsumexp is always defined stably as presented above.

\section{Benchmarked Tools}
\label{sec:tools}
We have chosen several well-known or promising AD tools (see~\tabref{tab:all}). The selection covers various languages and AD approaches as well as symbolic differentiation. The newest version of all the tools that was available in the period July-August 2015 was used. In addition, we give results for \emph{finite differences} and \emph{manual}, i.e., a hand-derived optimized implementation. 

The tools that have both forward and reverse mode are called with the one that is more suitable for the given objective function. Diffsharp in particular runs significantly slower in its default mode so it is called in its special forward and reverse modes for first-order derivatives. 

Tapenade offers differentiation of both Fortran and clean C code but we use it only with~C. Unfortunately, it does not support C fully and its source transformation occasionally produced non-compiling output, so we had to fix a few errors. 

From the chosen tools, we did not benchmark ADiGator because it generated syntactically incorrect code for GMM, clad as it did not have support for arrays, and ADIC2 as our attempts to compile it were unsuccessful. Consider that this supports the statement that it is more difficult to implement a source transformation \changecol{than an operator overloading tool}.

Adept, ADOL-C and Ceres are all operator overloading \Cpp{} tools. They all require a templated objective function as input so that it could be run with their custom types computing derivatives. \changecol{Ceres has a straightforward implementation of forward mode only. ADOL-C implements both forward and reverse modes using so called \emph{taping} which is basically a process of storing all calculations involving active variables. Importantly, the tape can be reused for successive computations assuming that certain conditions hold. Adept is based on a similar idea but it makes use of expression templates. That makes the taping process efficient enough so that it can be run for every computation without incurring any significant slowdown.}

MuPAD (called from MATLAB) optimizes code using common subexpression elimination and compiles it via \Cpp{} to MEX. Theano input needs to be written in a modified Python and is then compiled either into optimized Python or \Cpp{}. Theano is always ran in CPU mode to allow a fair comparison since all the tools use only CPU.

\begin{table}
\tbl{List of tools. OO: operator overloading, ST: source transformation: F: forward, R: reverse.}
{\begin{tabular}[l]{@{}|c|c|c|c|}
        \hline
        Language    & Tool          & Approach  & Mode \\ \hline
        \hline 
        \Cpp{}         &                       & Manual (by hand)          &          \\ \hline
        \Cpp{}         &                       & Finite differences        &          \\ \hline
        \Cpp{}         & Adept~\cite{Hogan14adept}             & OO        & F, R     \\ \hline
        \Cpp{}         & ADIC2~\cite{Narayanan10adic2}         & ST        & F, R       \\ \hline
        \Cpp{}         & ADOL-C~\cite{Walther12adolc}          & OO        & F, R     \\ \hline
        \Cpp{}         & Ceres Solver~\cite{ceres}             & OO        & F          \\ \hline
        \Cpp{}         & clad~\cite{clad}                      & ST via compiler & F    \\ \hline
        C\grey{/Fortran} & Tapenade~\cite{Hascoet13tapenade} & ST       & F, R      \\ \hline
        F\#         & DiffSharp~\cite{Baydin15survey}     & OO        & F, R     \\ \hline
        MATLAB      & ADiGator~\cite{adigator}              & ST via OO & F          \\ \hline
        MATLAB      & ADiMat~\cite{Bischof02adimat}         & OO via ST & F, R    \\ \hline
        MATLAB      & MuPAD~\cite{mupad}                    & Symbolic  &             \\ \hline
        Julia       & ForwardDiff.jl~\cite{julia-fwd-diff}  & OO        & F        \\ \hline
        Python      & Autograd~\cite{autograd}              & OO        & F        \\ \hline
        Python      & Theano~\cite{Bastien12theano} & Symbolic &   \\ \hline
\end{tabular}}
\label{tab:all}
\end{table}

\section{Automatic Differentiation in Computer Vision and Machine Learning}

Problems in computer vision and machine learning are often formulated as non-linear optimization. Some of these problems are neural network training, bundle adjustment, clustering or tracking, to name a few. Optimization algorithms typically require derivatives in the form of gradients, Jacobians, or Hessians. Therefore, AD methods can be applied in these fields. They can prove very useful, especially during prototyping, as the objective function may be changed as often as the programmer wishes without putting any effort into derivative-computation implementation and still get exact derivatives. Nonetheless, AD methods are still not widely known in the machine learning and computer vision community.

In the cases, where the community applies AD or AD-like techniques, specialized tools are typically employed instead of existing general AD implementations. 
This also motivates our benchmark to see how they compare. 
These specialized tools are Ceres~\cite{ceres}, Autograd~\cite{autograd}, and Theano~\cite{Bastien12theano}, for example. Ceres implements a simple forward mode AD in \Cpp{}, 
Autograd is a reverse mode implementation for Python, and 
Theano is a collection of symbolic and AD-like differentiation methods using its own syntax based on Python. 

Another related technique, used for training neural networks, is the backpropagation algorithm, essentially a special case of reverse-mode AD.
For a more comprehensive survey of AD in machine learning, see Baydin \etal~\cite{Baydin15survey}.

\section{Objective Functions}
\label{sec:ofun}
In this section, we present the three objective functions used for benchmarking AD tools. The functions are: log-likelihood of a Gaussian mixture model, bundle adjustment, and hand tracking. 

\subsection{Objective GMM: Gaussian Mixture Model Fitting}
\label{sec:gmm}
The Gaussian mixture model can be used in a wide range of applications. Consider
clustering,
deblurring of images~\cite{Zoran11gmm}
and speech recognition~\cite{Yu14automatic}
for instance.
The GMM has likelihood function
\begin{equation}
\begin{gathered}
\begin{aligned}
p(\mX;\vw,\mMu,\tSigma) = \prod_{i=1}^{N}{ \sum_{k=1}^K{w_k {|2 \pi \mSigma_k|}^{-\frac12}} \exp{\left(-\frac{1}{2} (\vx_i-\vmu_k)^\transpose \mSigma_k^{-1} (\vx_i - \vmu_k)\right)}} \end{aligned}\\ 
\text{s.t.} \sum_{k=1}^K{w_k} = 1 \ \text{and} \ \mSigma_k \ \text{is positive-definite} \ \forall k \in \{1,\dots,K\}
\end{gathered}
\label{eq:gmm1}
\end{equation}
where variables $\vx_i \in \real^D$ are data points, $w_k \in \real$ weights, $\vmu_k \in \real^{D}$ means, and $\mSigma_k \in \real^{D \times D}$ covariance matrices. \changecol{Function inputs $\mX, \vw, \mMu, $ and $ \tSigma$ are their concatenations as explained in \secref{sec:not}}. 

We  parametrize the positive-definite covariance matrices by the square roots of their inverses. We introduce variables $\vq_k \in \real^D$ and $\vl_k \in \real^{\frac{D(D-1)}{2}}$ and function $Q(\vq, \vl)$ which assembles a $D \times D$ lower triangular matrix in the following way
\begin{equation}
Q(\vq, \vl) = \bmat \exp(q_1) & 0 & \cdots & 0 \\
                    l_1 & \exp(q_2) & \cdots & 0 \\
                    \vdots & \vdots & \ddots & \vdots\\
                    l_{D-1} & l_{D-1 + D-2} & \cdots & \exp(q_D) \emat.
\end{equation}
from which we assemble 
$\mSigma^{-1} = Q(\vq,\vl) Q(\vq,\vl)^\transpose.$

\noindent 
Positive weights $w_k$ are parameterized by log-parameters~$\alpha_k \in \mathbb R$:
\begin{equation}
w_k = \frac{\exp(\alpha_k)}{\sum_{k'=1}^K{\exp(\alpha_{k'})}}.
\end{equation}

In addition, we include an Identity-Wishart prior over the covariances
\begin{equation}
    p(\tSigma) = 
\prod_{k=1}^K C(D,m)|\mSigma_k|^m \exp\left(-\frac12 \operatorname{trace}(\mSigma_k)\right) 
\end{equation}
where variable $m$ is a Wishart prior hyperparameter and $C$ is a function not dependent on independent variables.

The goal of GMM inference is to maximise the posterior probability of data given parameters, or equivalently to minimize the negative log posterior
\[
    L(\vw,\mMu,\tSigma; \mX) = -\log\bigl(p(\mX;\vw,\mMu,\tSigma) p(\tSigma)\bigr)
\]
Discarding function $C$ and simplifying using the described parametrization, the final function to be optimized looks like
\begin{align}
L(\valpha,\mMu,\mQ,\mL) = 
&\sum_{i=1}^N{\logsumexp\left(\left[\alpha_k + \operatorname{sum}(\vq_k) - \frac{1}{2}||Q(\vq_k,\vl_k)(\vx_i-\vmu_k)||^2\right]_{k=1}^K\right)} \nonumber\\
&- N \logsumexp\left(\left[\alpha_k\right]_{k=1}^K\right) \label{eq:gmm}\\
&+ \frac{1}{2} \sum_{k=1}^{K}{\left(||\exp{(\vq_k)}||^2 + ||\vl_k||^2\right)} - m \operatorname{sum}(\vq_k) \nonumber
\end{align}

We benchmark the AD tools on gradient computation of~\equref{eq:gmm}. The size of the gradient changes with $D$ and $K$, while $\valpha, \mMu, \mQ$ and $\mL$ are independent variables.

\changecol{Note that it is possible to implement the first line of \equref{eq:gmm} using large matrix operations provided that enough memory is available. This can significantly speed up some languages and tools (see \secref{sec:exp}). The main idea is to work with all the data at once instead of using an outer loop over the data. For instance, we can compute
\begin{equation}
Q(\vq_k,\vl_k)\left(\mX-\bmat \vmu_k & \vmu_k & \hdots & \vmu_k \emat \right)
\end{equation}}
at the cost of $O(ND)$ words of storage.

\subsection{Objective BA: Bundle Adjustment}
\label{sec:ba}
In computer vision, 3D reconstruction is a widely studied problem~\cite{Snavely-siggraph06,Agarwal-acm11}. Given a visual input (\eg images or video) observing the same scene, the goal is to reconstruct a 3D model of this scene. Even though the creation of 3D models can be a goal on its own, 3D reconstruction is necessary for a number of other applications such as localization~\cite{Sattler12-localization}, robot navigation~\cite{Davison07monoslam}, augmented reality or virtual reality~\cite{Shapira16reality}.

Consider so called sparse 3D reconstruction. In this problem, given only images we want to find 3D coordinates of some points observed in the images together with parameters of cameras for the images, \ie, where the cameras were in the world when images were taken. That can be done by various approaches but most of them run an optimization procedure called bundle adjustment (BA)~\cite{Triggs99ba,ceres}. This procedure optimizes  simultaneously all the parameters, \ie, all 3D point coordinates and parameters of cameras. We benchmark the AD tools by computing the Jacobian used in BA.

Let us first introduce the projection function for one camera and one point. Consider a weight $w \in \real$, a 3D point $\vx \in \real^3$ and a camera with parameters $\vp = [\vr;\vc;f;\vx_0;\vkappa] \in \real^{11}$, i.e., rotation, camera center, focal length, principal point and radial distortion. The point $\vx$ can be projected by the camera as
\begin{align}
\operatorname{project}(\vp,\vx) &= \operatorname{distort}(\vkappa,\operatorname{p2e}(\operatorname{rodrigues}(\vr,\vx-\vc)))f + \vx_0 
\end{align}   
where
\begin{align}   
\operatorname{distort}(\vkappa,\vu) &= \vu(1 + \kappa_1 ||\vu||^2 + \kappa_2 ||\vu||^4) \\
\operatorname{p2e}(\vx) &= \frac{\vx_{1:2}}{x_3} \\
\operatorname{rodrigues}(\vr,\vx) &= \vx \cos{\theta} + (\vv \times \vx) \sin{\theta} + \vv(\vv^\transpose\vx)(1 - \cos{\theta}), \quad \theta = ||\vr||, \vv = \frac{\vr}{||\vr||}
\label{eq:rodrigues}
\end{align}   
The observed image point is $\vm \in \real^2$ and the residual $\bm e$ concatenates its reprojection error~\cite{HZ04} and $w$'s regularizer
\begin{equation}
\label{eq:residual}
\ve = [w (\vm - \operatorname{project}(\vr,\vc,f,\vx_0,\vkappa,\vx))^\transpose;
1 - w^2]^\transpose
\end{equation}

\begin{figure}
    \centering
    \framebox{\includegraphics[width=0.7\textwidth]{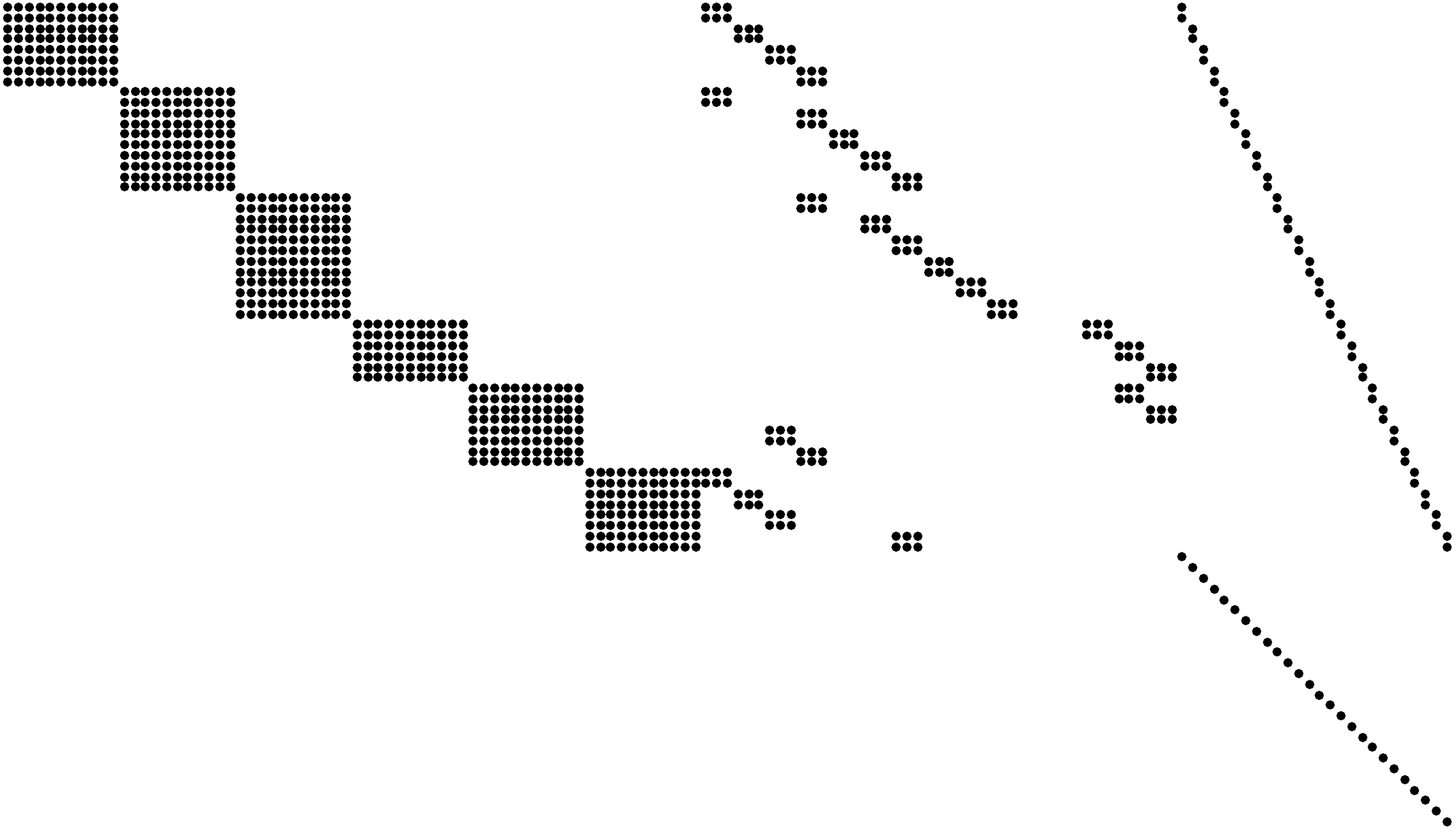}}
    \caption{Sparsity pattern of the Jacobian for an example instance of bundle adjustment. The first set of wider blocks corresponds to camera parameters, the middle set to 3D points, and the last set to weights.  Rows have been permuted so the ``weights-only'' rows appear after all the ``reprojection error'' rows.}
    \label{fig:ba-jacobian}
\end{figure}

The Jacobian of the whole system where multiple cameras observe multiple points has a special form. It has only 15 non-zero entries in every reprojection-error row and one non-zero in every weight-term row. See \figref{fig:ba-jacobian} for a visualization. \changecol{Importantly, every residual is independent of others. It is thus possible to compute small ($3 \times 15$) dense Jacobians corresponding to individual residuals by directly differentiating the residual function (see \equref{eq:residual}). Then, it is straightforward to distribute the entries across the final sparse Jacobian. Hence, AD tools are not required to support sparsity in any way in order to compute the Jacobian of this problem. This strategy is applied also in the popular optimizer Ceres, that is quite often used to solve BA problem in computer vision~\cite{ceres}.} Also note that because of the sparsity, the width of the Jacobian is not important and time complexity depends only on the number of observations.

\subsection{Objective HT: Hand Tracking}
In hand tracking~\cite{Taylor14hand}, we are given a model of a hand and a stream from a depth sensor. The goal is tracking a real hand observed by the depth sensor, \ie, fitting the model to the depth information. An application requiring hand tracking is remote control and interaction~\cite{Wood16hands}, for instance.

For benchmark purposes, let us consider only the optimization part of the hand tracking problem. We are given the hand model aligned to the previous frame. The model is a set of points $\mX \in \real^{3 \times M}$ and their triangulation, i.e., a collection of adjacent triangles, which make up a surface. The motion of the model is parametrized by the variable $\vp \in \real^{26}$. Then, we are given $N$ correspondences between the triangles and measured 3D points $\mY \in \real^{3 \times N}$ obtained from the current depth frame. The variable $\mU \in \real^{2 \times N}$ are barycentric coordinates defining exact spots of correspondence inside the triangles. Additionally, we are given weights $\mW \in \real^{22 \times M}$ defining which points lie on which parts of the hand (see the procedure below).

The variable $\vp$ contains 3 parameters for global translation, 3 for global rotation parametrized using angle-axis representation and 4 angles for every finger.

The procedure for computing the error for all measurements is based on linear blend skinning:
\begin{enumerate}
    \item Use the finger parameters to assemble 22 transformations $\tT \in \real^{4 \times 4 \times 22}$  corresponding to parts of hand. This operation first assembles individual independent relative transformations corresponding to joints using the Euler angles approach and then hierarchically combines them to the absolute transformations $\tT$. 
    \item Transform all model vertices by all transformations and weight by those that are relevant, \ie, 
    \begin{equation}
    \mZ = \sum_{i=1}^{22}{\mT_i \bmat \bar{\vx}_1^i & \bar{\vx}_2^i & \hdots & \bar{\vx}_M^i \emat} \in \real^{4 \times M}, \quad \bar{\vx}_j^i = w_{i,j} \bmat \vx_j \\ 1 \emat \in \real^{4}
    \end{equation}
    \item Apply global rotation and translation
    \begin{equation}
    \mV = \bmat \mR & \vt \emat \mZ \in \real^{3 \times M}
    \end{equation}
    Note that we can take $3 \times 4$ matrix because all $\mT_i$ are composed of rotation and translation only and weights for every point sum up to one. Therefore all $\vz_j$ have the fourth coordinate equal to one.
    \item Having transformed the hand model, find the exact correspondence spots inside the triangles. For $q$-th measurement corresponding to the triangle $(i,j,k)$:
    \begin{equation}
    \vy'_q = u_{q,1} \vv_i + u_{q,2} \vv_j + (1 - u_{q,1} - u_{q,2}) \vv_k
    \end{equation}
    which gives us $\mY' \in \real^{3 \times M}$.
    \item Finally, the errors for all points are simply $\mE = \mY - \mY'$.
\end{enumerate}

\begin{figure}
    \centering
    \framebox{\includegraphics[width=0.45\textwidth]{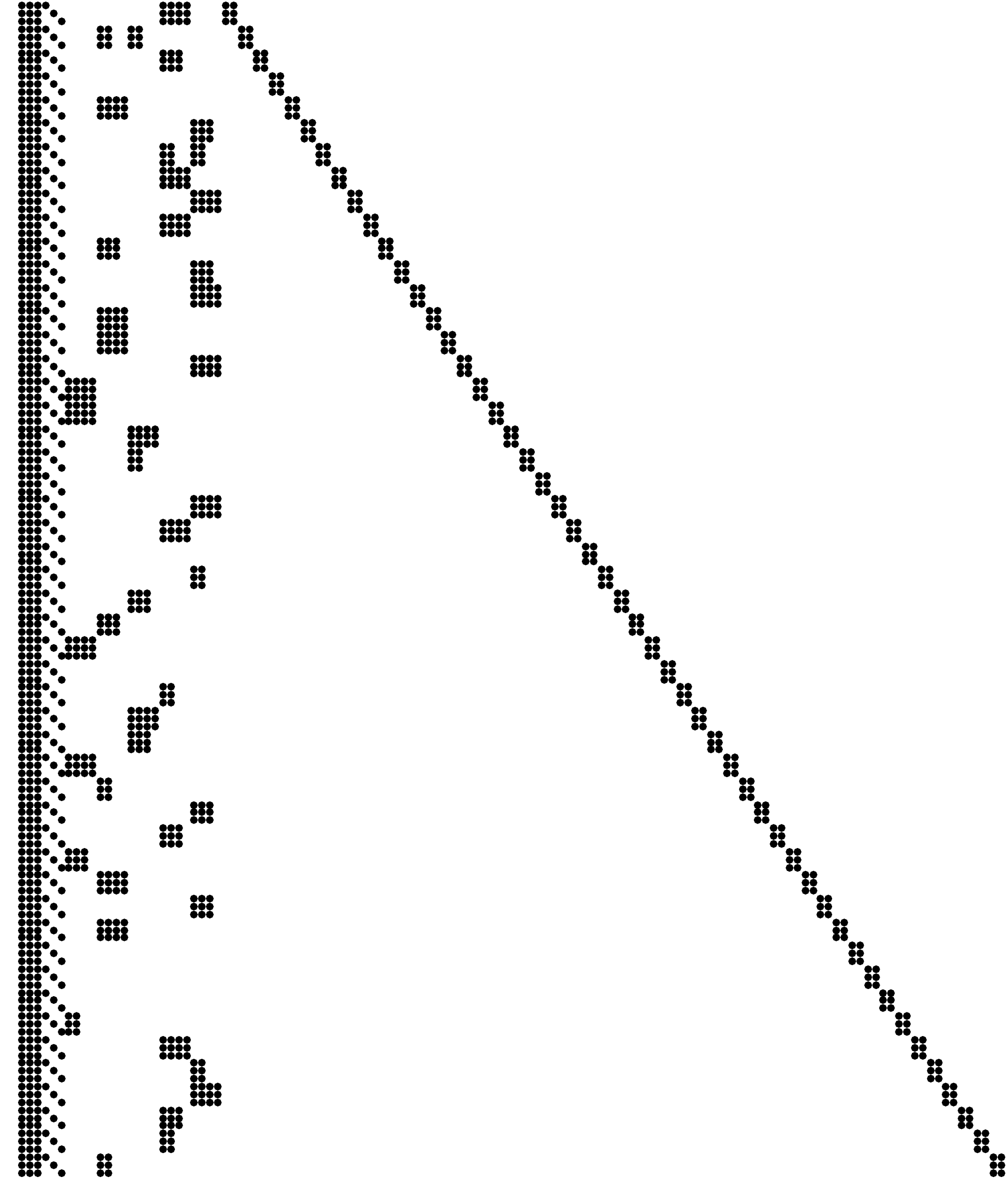}}
    \caption{Sparsity pattern of the Jacobian for an example instance of hand tracking. The left part corresponds to motion parameters and the diagonal part on the right to barycentric coordinates.}
    \label{fig:ht-jacobian}
\end{figure}

The independent variables are $\vp$ and $\mU$. We benchmark the Jacobian computation which has a special structure. It has a semi-dense mostly unstructured part composed of columns of $\vp$ and a sparse part corresponding to $\mU$, where every row has two non-zero entries. See \figref{fig:ht-jacobian} for a visualization. \changecol{In contrast to BA (see \secref{sec:ba}), it is not possible to compute individual blocks of the Jacobian independently. Therefore, sparsity has to be exploited differently for efficient Jacobian computation.}

\changecol{One has to create a seed matrix which defines the compression, feed it to an AD tool and decompress the resulting matrix. Having the sparsity pattern, it is possible to compute a seed matrix automatically using ColPack~\cite{colpack}, for example. Nevertheless, we propose to exploit the properties of the HT problem and design the seed matrix manually. 
The sparsity pattern of the (left) semi-dense part of the HT Jacobian can change in every iteration. Therefore, we propose to treat the left part as a dense Jacobian in order to avoid seed matrix computation cost. The number of columns of the left part is constant. Hence, the AD tools will always need the same number of function passes. 
The sparsity pattern of the (right) part is of a diagonal structure and does not change. It is straightforward to create a seed matrix which compresses the pattern of the right side into two columns.}

\section{Experiments}
\label{sec:exp}

To benchmark the AD tools, we first ran pre-processing routines (e.g. source transformation, symbolic differentiation, taping). All of the routines that need to be run only once for different data are not included in the runtimes that we provide. This is justified since a user of AD tools would typically run it only once on the objective before calling the differentiated function many times to optimize parameters.

The benchmarking is done on random data. The resulting runtimes are averaged over 1000 runs if one run is less than 5 seconds, over 100 runs if 5-30 seconds and over 10 runs if 30-120 seconds. Otherwise, the runtimes are not averaged. \changecol{The time limit for a single run is 40k seconds.} A single machine with a processor Intel(R) Xeon(R) CPU E5-1620 0 @ 3.60GHz, memory 32GB and OS Windows 10 64-bit was used for all the experiments.

We measure not only derivative-computation runtimes for every differentiation approach but also objective-computation runtimes for every language. Hence, we are able to show derivative runtimes for every approach relative to objective runtimes. Note that this measure attempts to minimize the dependence of results on individual languages. Throughout this section, relative runtimes refer to absolute derivative-computation runtimes divided by absolute objective-computation runtimes measured in a corresponding language. Special case are the symbolic differentiation tools Theano and MuPAD. For them, we record runtimes of objective computation which is optimized by their internal engines.

Note that visualizations and tables with results for absolute derivative-computation runtimes are provided in the supplementary material.

\subsection*{Experiment: Gaussian Mixture Model (GMM)}

\begin{figure}[t]
    \includegraphics[width=\textwidth]{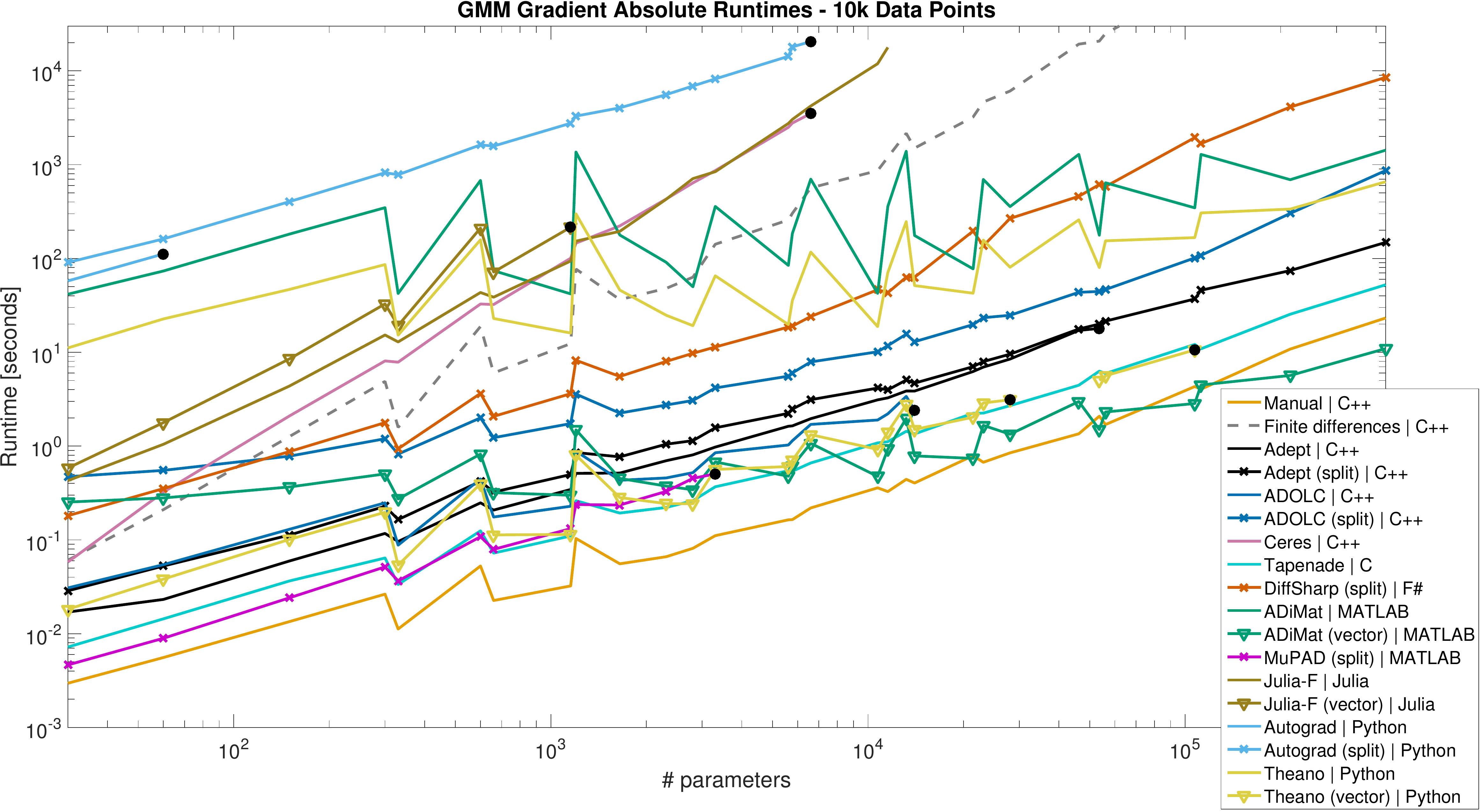}
    \caption{Absolute runtimes for GMM with 10k data points. Some of the tools were run with (split) or (vector) implementations (see \secref{sec:exp}). 
 The curve endings emphasized by the black dots symbolize that the tools crashed on bigger instances and those not emphasized did not finish in our time limit. 
 Note that both axes are log-scaled. Best viewed in color.}
    \label{fig:gmm-10k-abs}
\end{figure}

\figref{fig:gmm-10k-abs} shows gradient-computation runtimes for GMM with 10k data points. \changecol{Alternatively, see \tabref{tab:gmm-10k-abs} for a subset of the results.} We have noticed that tools Adept and ADOL-C do not handle bigger instances. 
DiffSharp and Autograd crash even for smaller instances. 
The biggest instance size ($D = 64$, $K = 200$) was taken from Zoran and Weiss~\cite{Zoran11gmm}. 
We help out these tools by \changecol{manually exploiting partial separability} by splitting the gradient computation into functions $f$, applied per datapoint, and $g$, the parts independent of datapoints
\begin{equation}
\nabla L(\valpha,\mMu,\mQ,\mL) = \sum_{i=1}^N{\nabla f(\vx_i;\valpha,\mMu,\mQ,\mL) + \nabla g(\valpha,\mQ,\mL)}
\end{equation}
which is symbolized by \emph{(split)} in the figures. This way, the tools are able to handle even the larger problem instances even though they require a lot of memory. 

Moreover, GMM allows for an opposite approach to (split), a vectorized implementation (denoted by \emph{(vector)}), where most necessary computations are done in one huge matrix multiplication (see \secref{sec:gmm}). We show this (vector) version with languages that are able to utilize it. Notice how Theano and ADiMat are boosted by (vector).

Note that MuPAD is the only tool having problems with compilation. It could not compile for larger problem sizes and it could take up to several hours to compile the others. Next, we point out that Ceres and Julia-ForwardDiff have forward mode only and \changecol{as can be seen, not having a reverse mode really puts them in a severe disadvantage, especially as the problem size grows}. The same holds for finite differences.

The relative runtimes for most of the tools fall in the range of two orders of magnitude. Interestingly, some tools perform very differently for different problem sizes. Taking ADiMat (vector), for instance, one can see that its relative runtime for the smallest problem size is much higher than for the largest one. We can only reason that it cannot utilize MATLAB's strength of matrix operations so much for the smaller data.

\changecolnew{Comparing standard and (split) versions of Adept, ADOL-C and Autograd, we observe a drop in runtime for all these tools when the (split) version is used. We argue that this is caused by multiple invocations of the taping process instead of just one. This claim is supported by the measured runtime difference between standard and (split) versions of ADOL-C and Adept. Both tools are written in \Cpp{} and use similar ideas but Adept employs efficient expression templates for taping. Hence, multiple invocations of the taping process do not incur a significant slowdown as opposed to ADOL-C.}

\changecol{We have also tried running the tools with 2.5M data points which is a number reported to be used in~\cite{Zoran11gmm}. With so many points, no tool could handle the biggest problem sizes without manual exploitation of partial separability. Implementations utilizing large matrix operations (denoted by (vector)) did not work at all as they need too much memory and cannot exploit partial separability by definition.}

\def\X{$\bullet$}

\begin{table}[t]
\tbl{Absolute runtimes for GMM with 10k data points. The bullet symbolizes that a tool crashed and no entry means that a tool did not finish in the time limit.}
{\begin{tabular}[l]{|l|l|llllllll|}
\hline
\multicolumn{2}{|l|}{\# parameters} & \num{3.00e+01} & \num{3.30e+02} & \num{1.20e+03} & \num{3.30e+03} & \num{1.07e+04} & \num{2.15e+04} & \num{5.36e+04} & \num{4.29e+05} \\ \hline \hline
Manual & C++ & \num{2.96e-03} & \num{1.12e-02} & \num{1.04e-01} & \num{1.11e-01} & \num{3.59e-01} & \num{7.90e-01} & \num{2.08e+00} & \num{2.32e+01} \\ \hline
Finite differences & C++ & \num{6.07e-02} & \num{1.58e+00} & \num{7.72e+01} & \num{1.42e+02} & \num{8.64e+02} & \num{3.23e+03} & \num{2.08e+04} & \\ \hline
Adept & C++ & \num{1.70e-02} & \num{9.61e-02} & \num{5.12e-01} & \num{9.76e-01} & \num{3.11e+00} & \num{6.24e+00} & \num{1.80e+01} & \multicolumn{1}{c|}{\X} \\ \hline
Adept (split) & C++ & \num{2.86e-02} & \num{1.65e-01} & \num{8.54e-01} & \num{1.57e+00} & \num{4.15e+00} & \num{7.03e+00} & \num{2.00e+01} & \num{1.48e+02} \\ \hline
ADOLC & C++ & \num{3.08e-02} & \num{8.79e-02} & \num{8.84e-01} & \num{8.49e-01} & \num{1.90e+00} & \multicolumn{1}{c}{\X} & \multicolumn{1}{c}{\X} & \multicolumn{1}{c|}{\X} \\ \hline
ADOLC (split) & C++ & \num{4.71e-01} & \num{8.22e-01} & \num{3.58e+00} & \num{4.17e+00} & \num{1.01e+01} & \num{1.97e+01} & \num{4.45e+01} & \num{8.66e+02} \\ \hline
Ceres & C++ & \num{5.80e-02} & \num{7.85e+00} & \num{1.46e+02} & \num{8.65e+02} & \multicolumn{1}{c}{\X} & \multicolumn{1}{c}{\X} & \multicolumn{1}{c}{\X} & \multicolumn{1}{c|}{\X} \\ \hline
Tapenade & C & \num{7.21e-03} & \num{3.35e-02} & \num{2.61e-01} & \num{3.68e-01} & \num{1.08e+00} & \num{2.24e+00} & \num{6.29e+00} & \num{5.25e+01} \\ \hline
DiffSharp (split) & F\# & \num{1.81e-01} & \num{9.36e-01} & \num{8.22e+00} & \num{1.14e+01} & \num{4.64e+01} & \num{1.96e+02} & \num{6.13e+02} & \num{8.53e+03} \\ \hline
ADiMat & MATLAB & \num{4.16e+01} & \num{4.24e+01} & \num{1.36e+03} & \num{3.59e+02} & \num{4.25e+01} & \num{7.75e+01} & \num{1.77e+02} & \num{1.43e+03} \\ \hline
ADiMat (vector) & MATLAB & \num{2.53e-01} & \num{2.73e-01} & \num{1.49e+00} & \num{6.77e-01} & \num{4.75e-01} & \num{7.39e-01} & \num{1.50e+00} & \num{1.10e+01} \\ \hline
MuPAD (split) & MATLAB & \num{4.64e-03} & \num{3.66e-02} & \num{2.38e-01} & \num{5.06e-01} & \multicolumn{1}{c}{\X} & \multicolumn{1}{c}{\X} & \multicolumn{1}{c}{\X} & \multicolumn{1}{c|}{\X} \\ \hline
Julia-F & Julia & \num{4.28e-01} & \num{1.29e+01} & \num{1.53e+02} & \num{8.42e+02} & \num{1.19e+04} & & & \\ \hline
Julia-F (vector) & Julia & \num{5.83e-01} & \num{1.93e+01} & \multicolumn{1}{c}{\X} & \multicolumn{1}{c}{\X} & \multicolumn{1}{c}{\X} & \multicolumn{1}{c}{\X} & \multicolumn{1}{c}{\X} & \multicolumn{1}{c|}{\X} \\ \hline
Autograd & Python & \num{5.76e+01} & \multicolumn{1}{c}{\X} & \multicolumn{1}{c}{\X} & \multicolumn{1}{c}{\X} & \multicolumn{1}{c}{\X} & \multicolumn{1}{c}{\X} & \multicolumn{1}{c}{\X} & \multicolumn{1}{c|}{\X} \\ \hline
Autograd (split) & Python & \num{9.07e+01} & \num{7.82e+02} & \num{3.30e+03} & \num{8.22e+03} & \multicolumn{1}{c}{\X} & \multicolumn{1}{c}{\X} & \multicolumn{1}{c}{\X} & \multicolumn{1}{c|}{\X} \\ \hline
Theano & Python & \num{1.11e+01} & \num{1.52e+01} & \num{2.99e+02} & \num{6.53e+01} & \num{1.88e+01} & \num{4.26e+01} & \num{8.00e+01} & \num{6.58e+02} \\ \hline
Theano (vector) & Python & \num{1.82e-02} & \num{5.38e-02} & \num{8.01e-01} & \num{5.64e-01} & \num{9.22e-01} & \num{2.03e+00} & \num{5.03e+00} & \multicolumn{1}{c|}{\X} \\ \hline
\end{tabular}}
\label{tab:gmm-10k-abs}
\end{table}

\begin{table}[t]
\def\X{$\bullet$}
\tbl{Absolute runtimes for GMM with 2.5M data points. The bullet symbolizes that a tool crashed and no entry means that a tool did not finish in the time limit. Only tools that could compute at least one problem instance are shown.}
{\begin{tabular}[l]{|l|l|llllllll|}
\hline
\multicolumn{2}{|l|}{\# parameters} & \num{3.00e+01} & \num{3.30e+02} & \num{1.20e+03} & \num{3.30e+03} & \num{1.07e+04} & \num{2.15e+04} & \num{5.36e+04} & \num{4.29e+05} \\ \hline \hline
Manual & C++ & \num{8.43e-01} & \num{3.29e+00} & \num{2.85e+01} & \num{3.01e+01} & \num{7.65e+01} & \num{3.80e+02} & \num{3.89e+02} & \num{6.16e+03} \\ \hline
Finite differences & C++ & \num{1.25e+01} & \num{3.54e+02} & \num{1.76e+04} & \num{3.31e+04} & & & & \\ \hline
Adept & C++ & \num{3.61e+00} & \multicolumn{1}{c}{\X} & \multicolumn{1}{c}{\X} & \multicolumn{1}{c}{\X} & \multicolumn{1}{c}{\X} & \multicolumn{1}{c}{\X} & \multicolumn{1}{c}{\X} & \multicolumn{1}{c|}{\X} \\ \hline
Adept (split) & C++ & \num{5.32e+00} & \num{3.50e+01} & \num{1.66e+02} & \num{3.72e+02} & \num{7.86e+02} & \num{2.31e+03} & \num{4.09e+03} & \num{3.99e+04} \\ \hline
ADOLC (split) & C++ & \num{9.83e+01} & \num{1.77e+02} & \num{7.91e+02} & \num{9.88e+02} & \num{2.32e+03} & \num{4.83e+03} & \num{1.04e+04} & \\ \hline
Ceres & C++ & \num{1.59e+01} & \num{2.27e+03} & \num{3.26e+04} & & & & & \\ \hline
Tapenade & C & \num{1.60e+00} & \num{8.58e+00} & \num{6.68e+01} & \num{8.56e+01} & \multicolumn{1}{c}{\X} & \multicolumn{1}{c}{\X} & \multicolumn{1}{c}{\X} & \multicolumn{1}{c|}{\X} \\ \hline
Tapenade (split) & C & \num{3.92e+00} & \num{1.33e+01} & \num{7.97e+01} & \num{1.09e+02} & \num{2.68e+02} & \num{9.59e+02} & \num{1.32e+03} & \num{1.59e+04} \\ \hline
DiffSharp (split) & F\# & \num{4.35e+01} & \num{2.44e+02} & \num{1.94e+03} & \num{3.34e+03} & \num{3.19e+04} & & & \\ \hline
MuPAD (split) & MATLAB & \num{1.45e+00} & \num{1.09e+01} & \multicolumn{1}{c}{\X} & \multicolumn{1}{c}{\X} & \multicolumn{1}{c}{\X} & \multicolumn{1}{c}{\X} & \multicolumn{1}{c}{\X} & \multicolumn{1}{c|}{\X} \\ \hline
Julia-F & Julia & \num{9.86e+01} & \num{2.59e+03} & & & & & & \\ \hline
Julia-F (vector) & Julia & \num{1.03e+03} & \multicolumn{1}{c}{\X} & \multicolumn{1}{c}{\X} & \multicolumn{1}{c}{\X} & \multicolumn{1}{c}{\X} & \multicolumn{1}{c}{\X} & \multicolumn{1}{c}{\X} & \multicolumn{1}{c|}{\X} \\ \hline
Autograd (split) & Python & \num{2.35e+04} & & & & & & & \\ \hline
Theano & Python & \num{3.23e+03} & \num{2.79e+03} & \multicolumn{1}{c}{\X} & \multicolumn{1}{c}{\X} & \multicolumn{1}{c}{\X} & \multicolumn{1}{c}{\X} & \multicolumn{1}{c}{\X} & \multicolumn{1}{c|}{\X} \\ \hline
Theano (vector) & Python & \num{5.48e+00} & \multicolumn{1}{c}{\X} & \multicolumn{1}{c}{\X} & \multicolumn{1}{c}{\X} & \multicolumn{1}{c}{\X} & \multicolumn{1}{c}{\X} & \multicolumn{1}{c}{\X} & \multicolumn{1}{c|}{\X} \\ \hline
\end{tabular}}
\label{tab:gmm-2.5M}
\end{table}

\begin{figure}[t]
    \includegraphics[width=\textwidth]{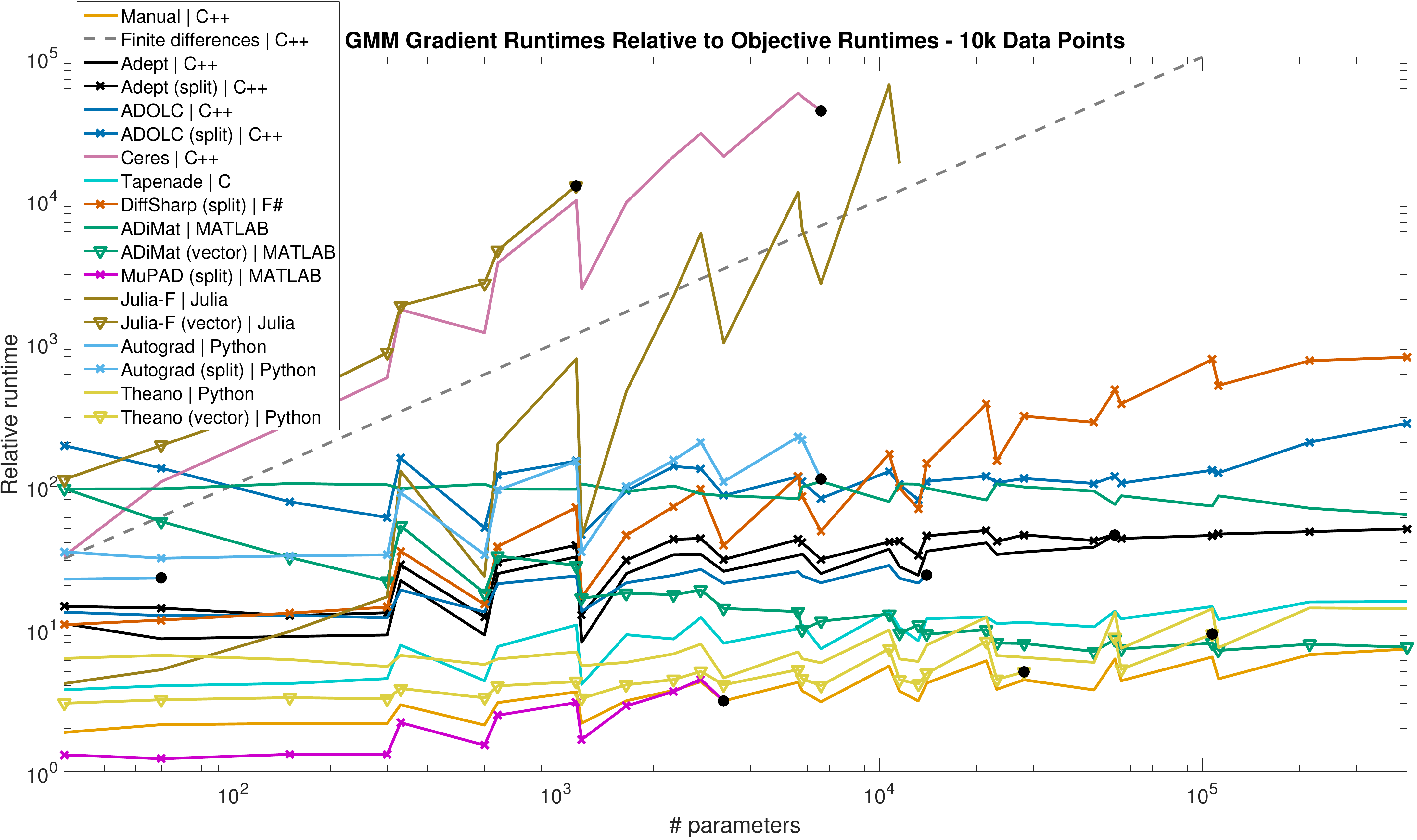}
    \caption{Relative runtimes for GMM with 10k data points. Some of the tools were run with (split) or (vector) implementations (see \secref{sec:exp}). 
  The curve endings emphasized by the black dots symbolize that the tools crashed on bigger instances and those not emphasized did not finish in our time limit. 
  Note that both axes are log-scaled. Best viewed in color.}
    \label{fig:gmm-10k-rel}
\end{figure}

\begin{figure}[t]
    \includegraphics[width=\textwidth]{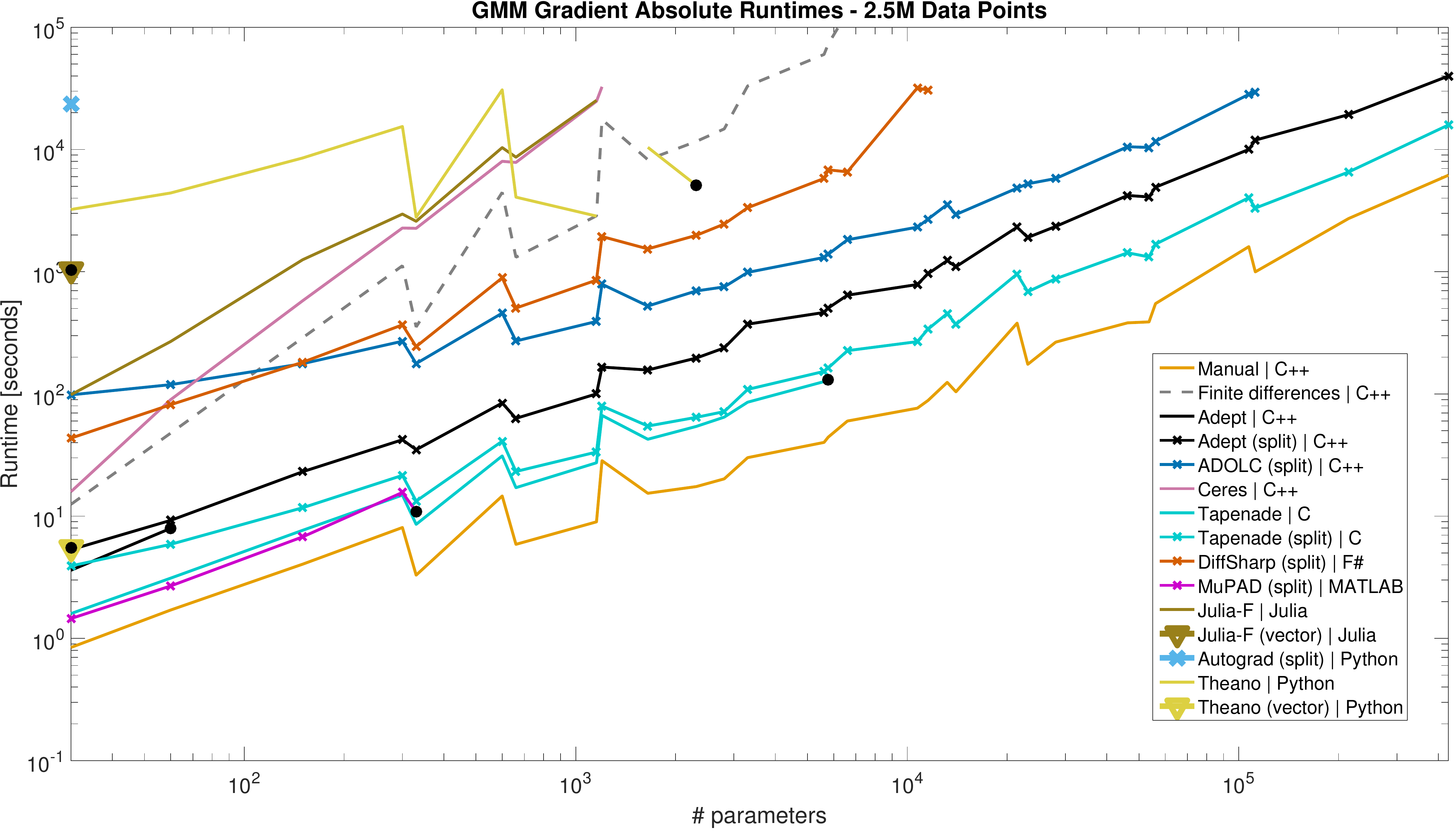}
    \caption{Absolute runtimes for GMM with 2.5M data points. Some of the tools were run with (split) or (vector) implementations (see \secref{sec:exp}). 
  The curve endings emphasized by the black dots symbolize that the tools crashed on bigger instances and those not emphasized did not finish in our time limit. 
  Only tools that could compute at least one problem instance are shown.
  Note that both axes are log-scaled. Best viewed in color.}
    \label{fig:gmm-2.5M}
\end{figure}

\clearpage

\subsection*{Experiment: Bundle Adjustment (BA)}

\begin{figure}[t]
    \includegraphics[width=\textwidth]{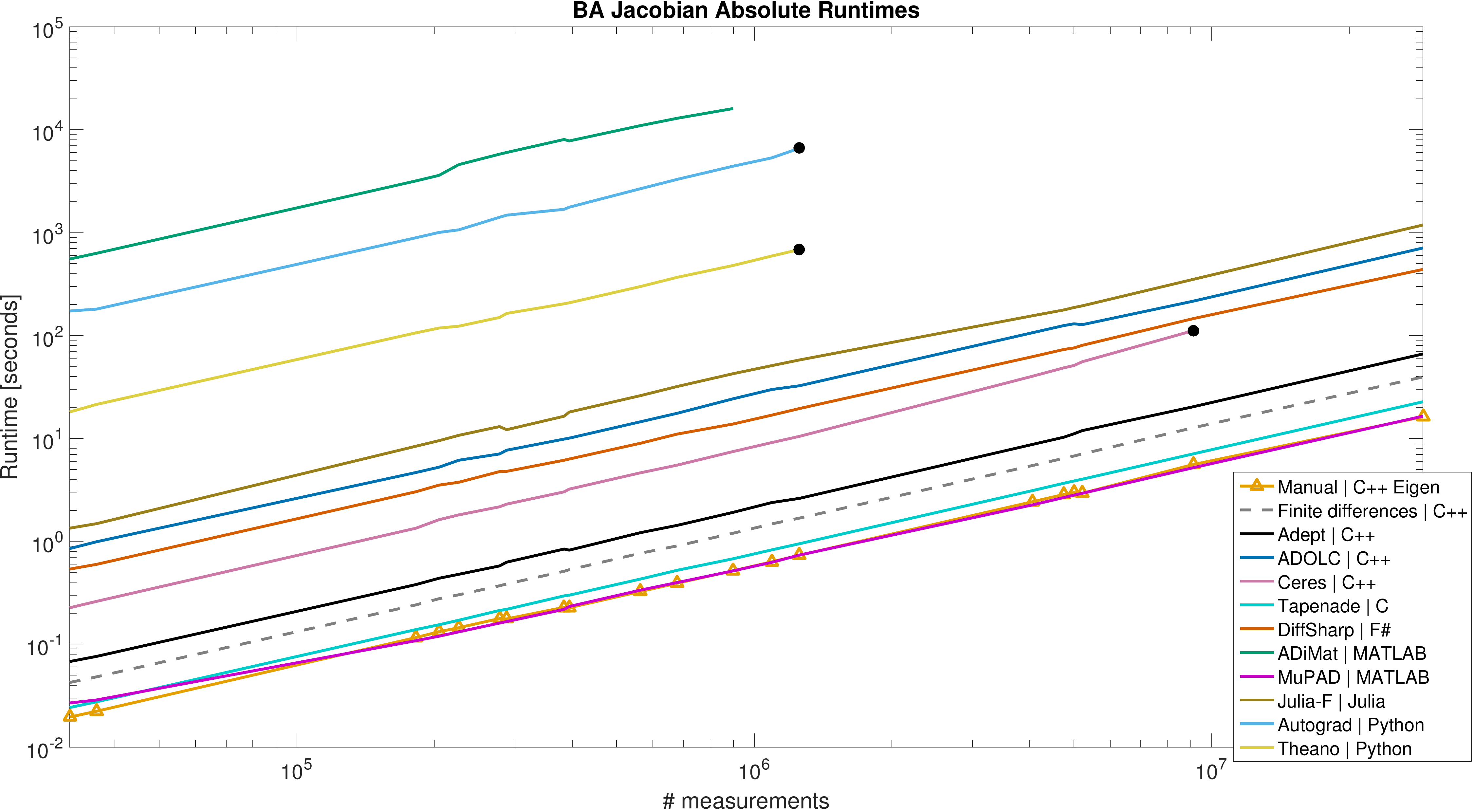}
    \caption{Absolute runtimes for BA. Note that Eigen matrix library~\cite{eigen} was utilized for implementing hand-derived derivatives. 
  The curve endings emphasized by the black dots symbolize that the tools crashed on bigger instances and those not emphasized did not finish in our time limit. 
  Note that both axes are log-scaled. Best viewed in color.}
    \label{fig:ba-abs}
\end{figure}

Next, we show Jacobian computation runtimes for BA in \figref{fig:ba-abs} \changecol{and \tabref{tab:ba-abs}}. We have chosen various problem sizes ranging from 21 cameras, 11k 3D points and 36k observations to 14k cameras, 4M 3D points, 29M observations. The problem sizes are samples of real-world dataset sizes~\cite{Agarwal10ba}. 

The more suitable mode for BA is reverse (see \secref{sec:ba}). Nevertheless, by comparing Ceres and ADOL-C, for instance, we can deduce that choosing either forward or reverse mode does not have so large significance in this case. ADiMat and Theano give inferior absolute runtimes as opposed to GMM with 10k data points, where they could utilize vectorization in the large matrix multiplication. Nevertheless, their relative runtimes are comparable to the other tools. Further notice that MuPAD is as good as manual implementation of the derivative computation. The reason for that is the use of common subexpression elimination and compilation into \Cpp{}.

\begin{table}[b]
\def\X{$\bullet$}
\tbl{Absolute runtimes for BA. Note that Eigen matrix library~\cite{eigen} was utilized for implementing hand-derived derivatives. The bullet symbolizes that a tool crashed and no entry means that a tool did not finish in the time limit. }
{\begin{tabular}[l]{|l|l|llllllll|}
\hline
\multicolumn{2}{|l|}{\# measurements} & \num{3.18e+04} & \num{2.04e+05} & \num{2.87e+05} & \num{5.64e+05} & \num{1.09e+06} & \num{4.75e+06} & \num{9.13e+06} & \num{2.90e+07} \\ \hline \hline
Manual & C++ & \num{1.96e-02} & \num{1.32e-01} & \num{1.76e-01} & \num{3.26e-01} & \num{6.32e-01} & \num{2.85e+00} & \num{5.58e+00} & \num{1.62e+01} \\ \hline
Finite differences & C++ & \num{4.25e-02} & \num{2.77e-01} & \num{3.85e-01} & \num{7.66e-01} & \num{1.48e+00} & \num{6.41e+00} & \num{1.27e+01} & \num{3.96e+01} \\ \hline
Adept & C++ & \num{6.79e-02} & \num{4.38e-01} & \num{6.28e-01} & \num{1.21e+00} & \num{2.38e+00} & \num{1.03e+01} & \num{2.03e+01} & \num{6.63e+01} \\ \hline
ADOLC & C++ & \num{8.50e-01} & \num{5.25e+00} & \num{7.68e+00} & \num{1.45e+01} & \num{2.99e+01} & \num{1.25e+02} & \num{2.16e+02} & \num{7.09e+02} \\ \hline
Ceres & C++ & \num{2.26e-01} & \num{1.62e+00} & \num{2.30e+00} & \num{4.63e+00} & \num{9.11e+00} & \num{4.85e+01} & \num{1.12e+02} & \multicolumn{1}{c|}{\X} \\ \hline
Tapenade & C & \num{2.43e-02} & \num{1.55e-01} & \num{2.18e-01} & \num{4.30e-01} & \num{8.26e-01} & \num{3.67e+00} & \num{7.09e+00} & \num{2.27e+01} \\ \hline
DiffSharp & F\# & \num{5.37e-01} & \num{3.52e+00} & \num{4.79e+00} & \num{8.98e+00} & \num{1.68e+01} & \num{7.32e+01} & \num{1.46e+02} & \num{4.39e+02} \\ \hline
ADiMat & MATLAB & \num{5.54e+02} & \num{3.60e+03} & \num{6.01e+03} & \num{1.10e+04} & & & & \\ \hline
MuPAD & MATLAB & \num{2.69e-02} & \num{1.20e-01} & \num{1.66e-01} & \num{3.36e-01} & \num{6.25e-01} & \num{2.66e+00} & \num{5.20e+00} & \num{1.65e+01} \\ \hline
Julia-F & Julia & \num{1.34e+00} & \num{9.51e+00} & \num{1.22e+01} & \num{2.61e+01} & \num{5.10e+01} & \num{1.77e+02} & \num{3.52e+02} & \num{1.19e+03} \\ \hline
Autograd & Python & \num{1.73e+02} & \num{1.00e+03} & \num{1.48e+03} & \num{2.67e+03} & \num{5.32e+03} & \multicolumn{1}{c}{\X} & \multicolumn{1}{c}{\X} & \multicolumn{1}{c|}{\X} \\ \hline
Theano & Python & \num{1.81e+01} & \num{1.18e+02} & \num{1.64e+02} & \num{3.00e+02} & \num{5.92e+02} & \multicolumn{1}{c}{\X} & \multicolumn{1}{c}{\X} & \multicolumn{1}{c|}{\X} \\ \hline
\end{tabular}}
\label{tab:ba-abs}
\end{table}

\begin{figure}[t]
    \includegraphics[width=\textwidth]{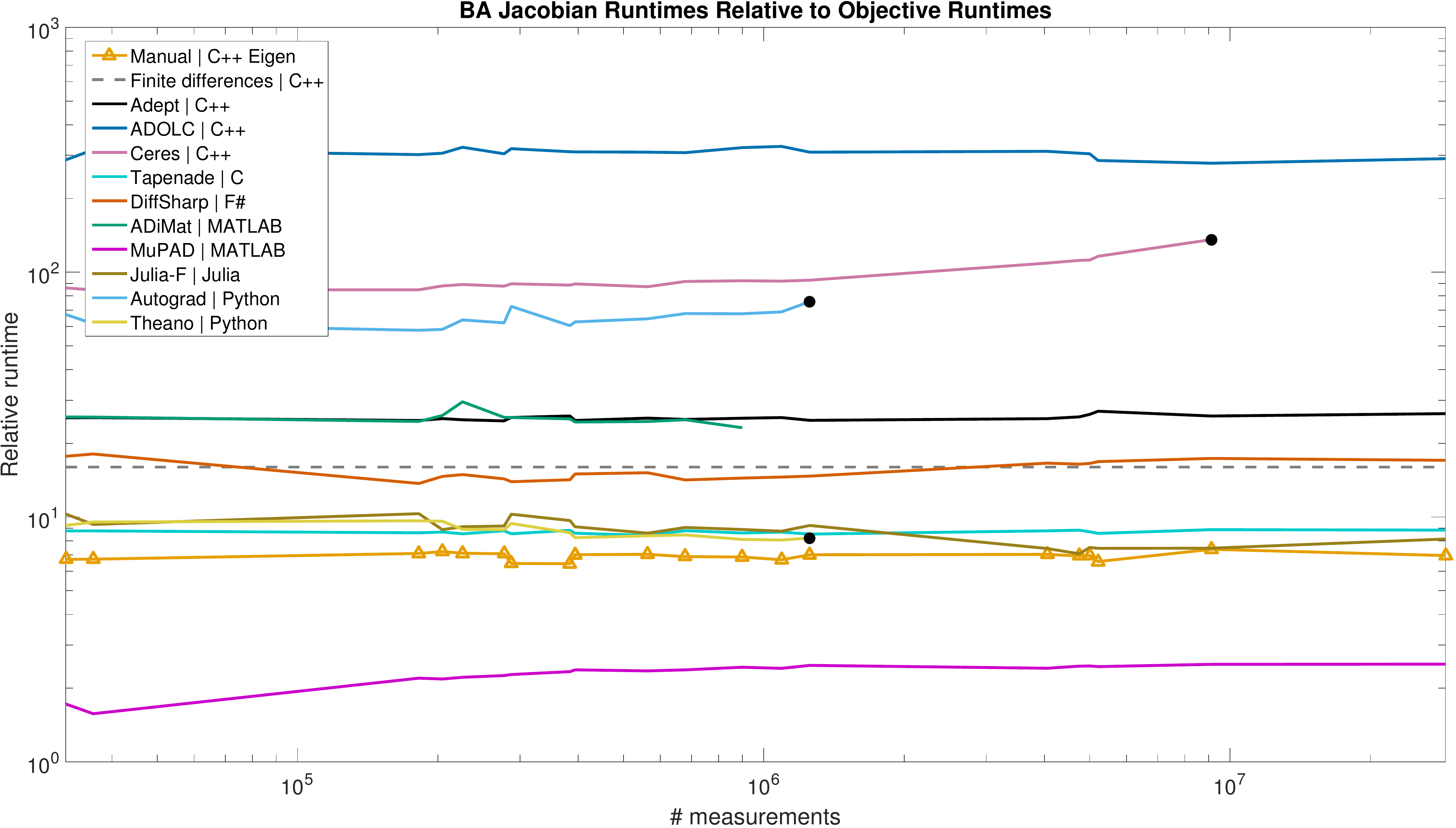}
    \caption{Relative runtimes for BA. Note that Eigen matrix library~\cite{eigen} was utilized for implementing hand-derived derivatives. 
  The curve endings emphasized by the black dots symbolize that the tools crashed on bigger instances and those not emphasized did not finish in our time limit. 
  Note that both axes are log-scaled. Best viewed in color.}
    \label{fig:ba-rel}
\end{figure}

\clearpage

\subsection*{Experiment: Hand Tracking (HT)}

\begin{figure}[t]
    \includegraphics[width=\textwidth]{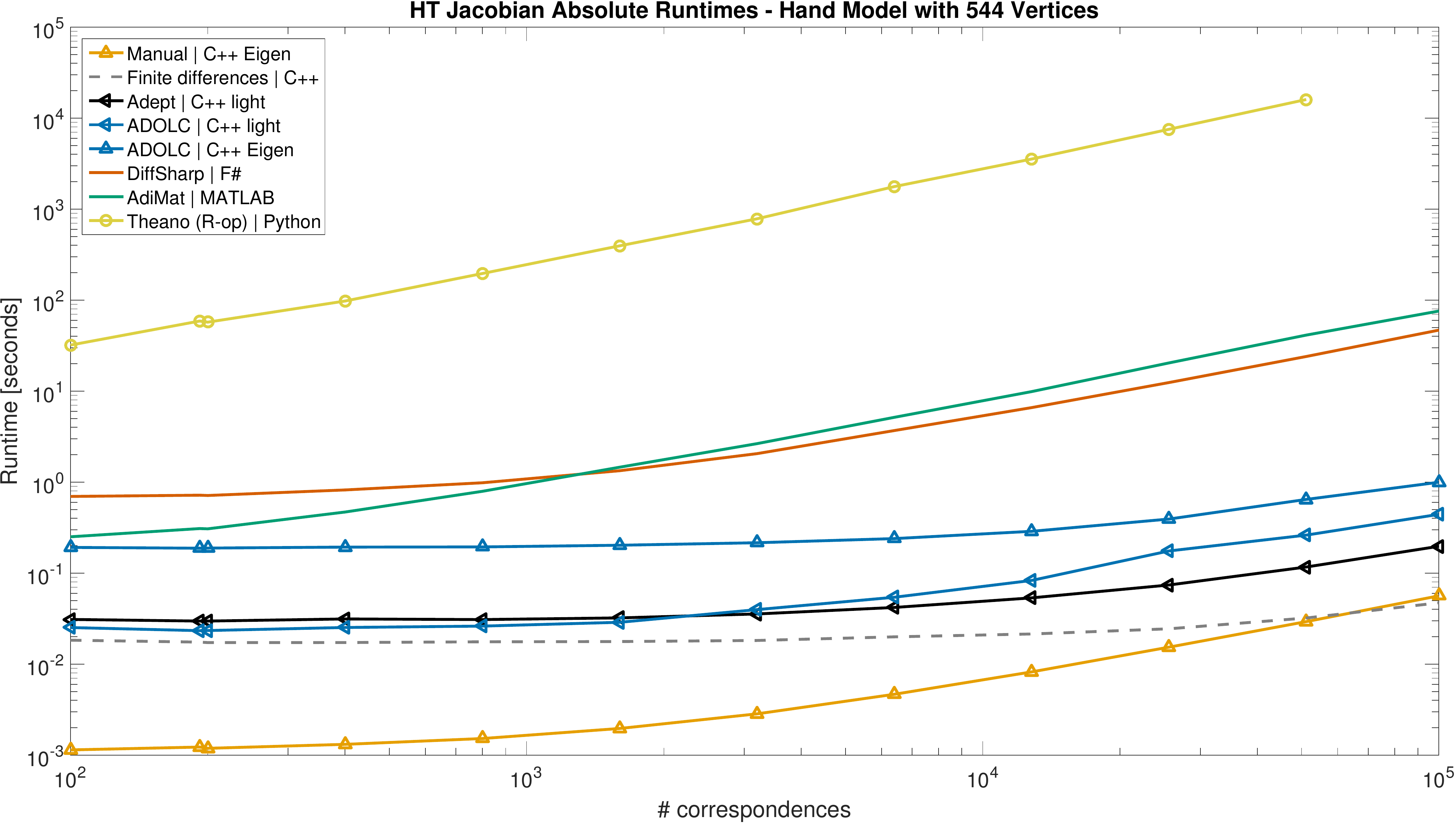}
    \caption{Absolute runtimes for HT with the smaller hand model.
  Only some tools were benchmarked (see \secref{sec:exp}).
  Theano did not finish in our time limit for the largest number of correspondences. 
  Note that both axes are log-scaled. Best viewed in color.}
    \label{fig:ht-small-abs}

    \includegraphics[width=\textwidth]{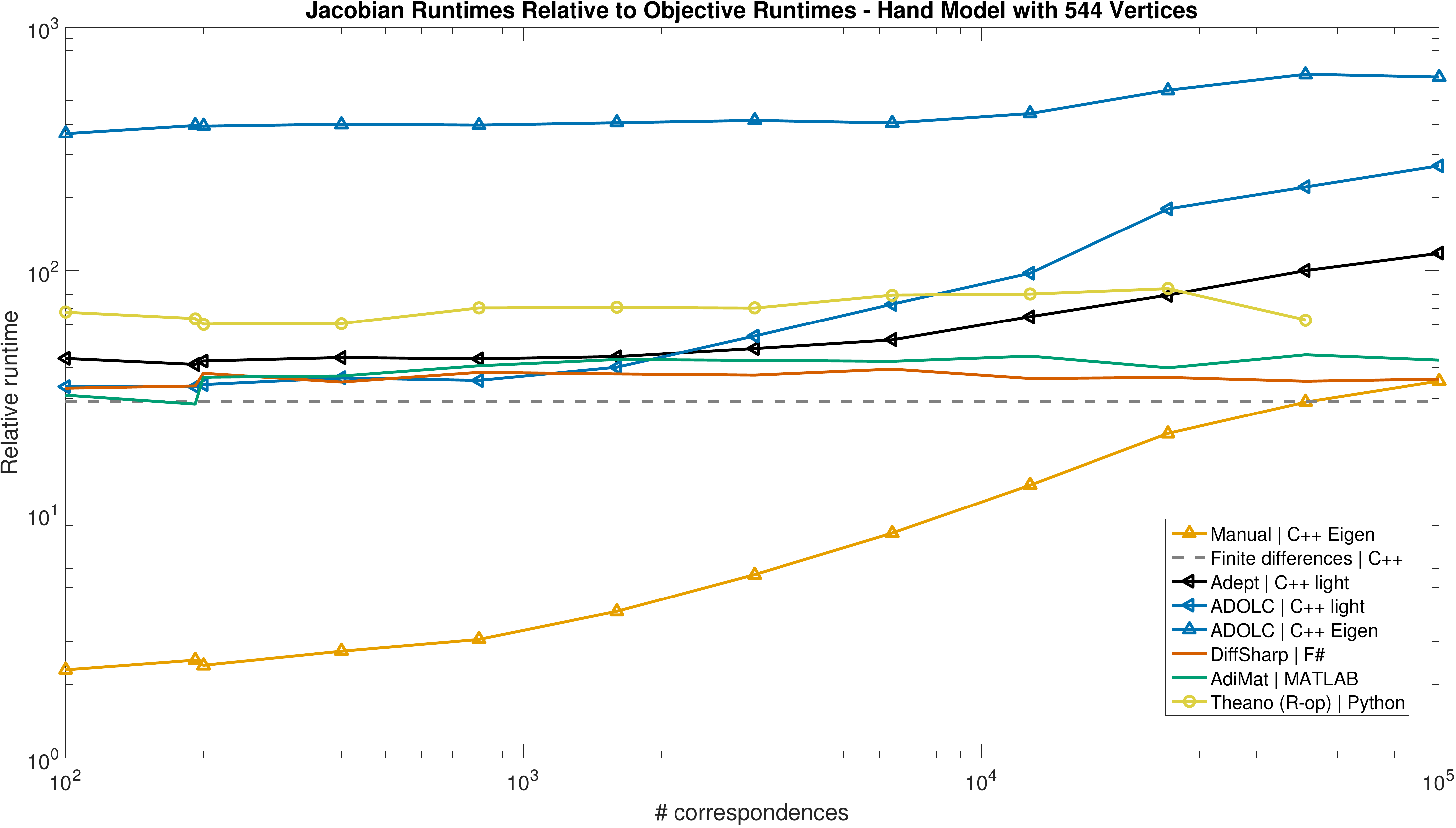}
    \caption{Relative runtimes for HT with the smaller hand model.
  Only some tools were benchmarked (see \secref{sec:exp}).
  Theano did not finish in our time limit for the largest number of correspondences. 
  Note that both axes are log-scaled. Best viewed in color.}
    \label{fig:ht-small-rel}
\end{figure}

\begin{figure}[t]
    \includegraphics[width=\textwidth]{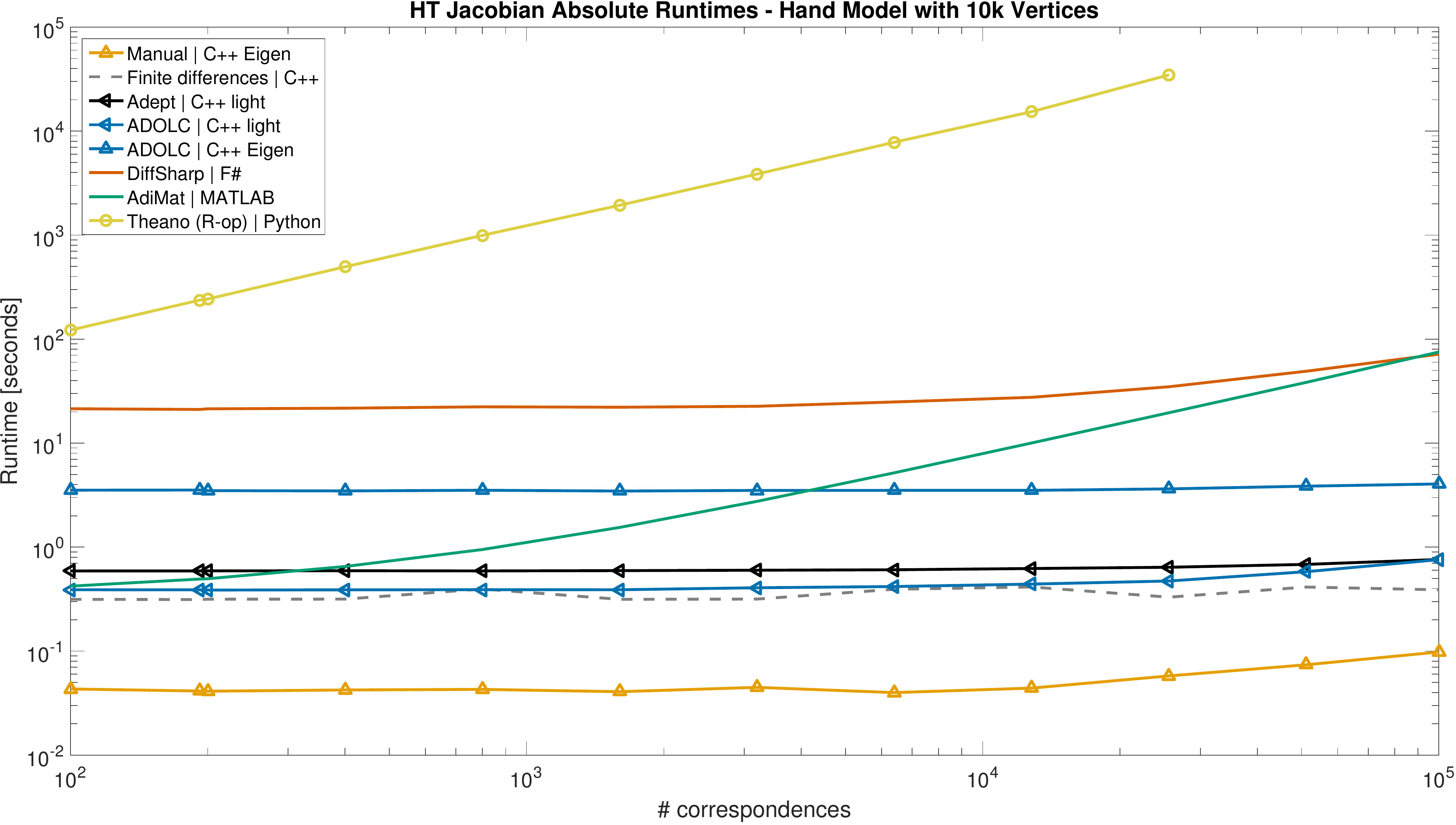}
    \caption{Absolute runtimes for HT with the larger hand model.
  Only some tools were benchmarked (see \secref{sec:exp}).
  Theano did not finish in our time limit for the two largest numbers of correspondences. 
  Note that both axes are log-scaled. Best viewed in color.}
    \label{fig:ht-big-abs}

    \includegraphics[width=\textwidth]{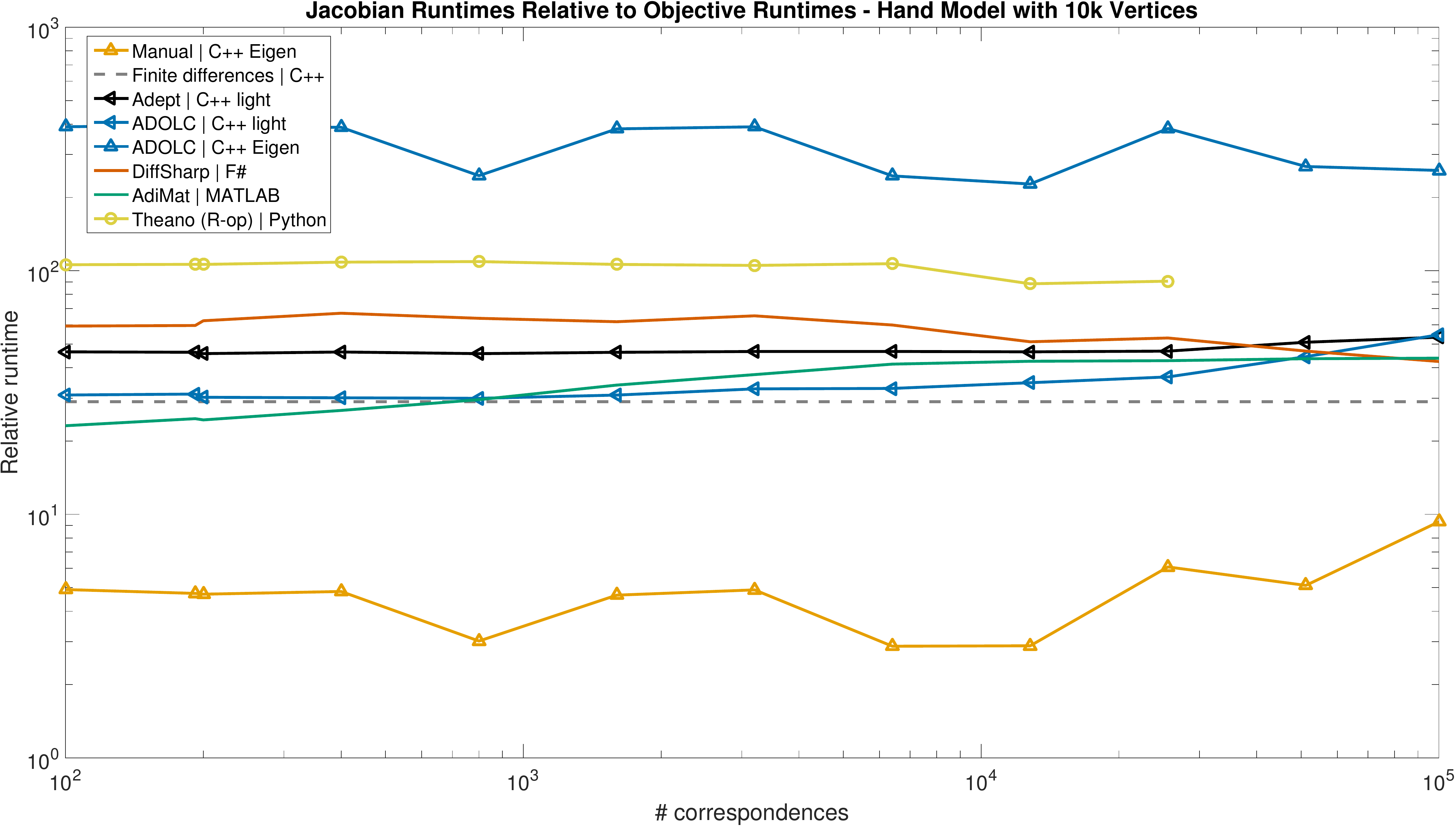}
    \caption{Relative runtimes for HT with the larger hand model.
  Only some tools were benchmarked (see \secref{sec:exp}).
  Theano did not finish in our time limit for the two largest numbers of correspondences. 
  Note that both axes are log-scaled. Best viewed in color.}
    \label{fig:ht-big-rel}
\end{figure}

For HT, we have chosen a small model size suitable for a real-time application and a larger one which would be typically run offline. The small instance has 544 points on the hand model and 192 correspondences whereas the big one has 10k points and 100k correspondences. We give Jacobian-computation runtimes for varying number of correspondences for the small model in \figref{fig:ht-small-rel}. The results for the large model are visualized in \figref{fig:ht-big-rel}.

Several tools were not benchmarked on HT. MuPAD had compilation issues. \changecol{Julia and Ceres did not allow for use of a custom seed matrix.} 
Tapenade was not benchmarked because the objective contains a lot of matrix operations which would have to be implemented in clean C. 
That is surely possible but consider that Tapenade does not support C fully and manually fixing the generated errors would require a significant effort. 
Finally, Autograd was not benchmarked as it implements only reverse mode.

The objective function in \Cpp{} was implemented in two different ways. One is using the Eigen matrix library~\cite{eigen} and the other using a custom lightweight matrix class (denoted in the figures by \emph{light}). \changecol{We use the custom class only with Adept because it is not compatible with Eigen and ADOL-C because Eigen is not optimized for the adouble class of ADOL-C. As can be seen, ADOL-C gives almost an order of magnitude worse results for the Eigen implementation than for the custom matrix class in terms of relative runtime of Jacobian-computation.} Further note that Theano's AD-like mode called R-op is used for HT. Its standard symbolic mode would not handle sparsity.

\clearpage

\section{Conclusion}
\label{sec:conc}

First, we have introduced automatic differentiation and chosen several tools for computing derivatives to be benchmarked. Second, we have pointed out the significance of derivatives in machine learning and computer vision and subsequently described three real-world objective functions from these areas. Then, we have provided relative runtimes for computing derivatives.

\changecol{We have seen that the relative runtimes of derivative computation range through three orders of magnitude. The relative runtime minimizes the effect of a programming language.}
Nevertheless, the runtime will still depend on programmer skill, and familiarity with the tools, so we have made open source all our materials\footnote{https://github.com/awf/autodiff}, in order that others may improve on our efforts.    However, we contend that this paper presents an important datapoint: a skilled programmer devoting roughly a week to each tool produced the timings above.  For many projects, these will represent typical results achieved before a tool is selected.

We conclude that there are useful tools in most languages but there is also still some space for improvement. Availability of various features proves to be crucial for the success and efficiency of algorithmic differentiation. Important features for our objectives include \changecol{ability to use a custom seed matrix}, support of matrix libraries, \changecol{partial separability detection}, and memory optimizations for big problem instances. Moreover, using the more suitable mode (forward or reverse) can really make a difference, especially for large problems. Therefore, availability of both modes in the AD tools is an advantage.
Importantly, note that we benchmarked only computation of the first-order derivatives and some tools do not support higher-order derivatives.

\section*{Acknowledgements}

This work was done while the first author was an intern at Microsoft Research.  
We thank Jonathan Taylor for an example implementation of a hand tracking function in Python.

\section*{Funding}

Zuzana Kukelova was supported by The Czech Science Foundation Project GACR P103/12/G084.

\bibliographystyle{gOMS}
\bibliography{biblio}

\end{document}